\documentclass[aps,preprintnumbers, superscriptaddress, amsmath, showpacs,]{revtex4}
\usepackage{epsfig}

\usepackage{graphicx}
\usepackage{dcolumn}
\usepackage{bm}

\begin{document}


\title{Equation of State in a Strongly Interacting Relativistic System}

\author{Efrain J. Ferrer  and   Jason P. Keith\\\textit{Department of Physics, University of Texas at El Paso, El Paso, Texas
79968, USA}}

\date{\today}

\begin{abstract}

We study the evolution of the equation of state of a strongly interacting quark system as a function of the diquark interaction strength. We show that for the system to avoid collapsing into a pressureless Boson gas at sufficiently strong diquark coupling strength, the diquark-diquark repulsion has to be self-consistently taken into account. In particular, we find that the tendency at zero temperature of the strongly interacting diquark gas to condense into the system ground state is compensated by the repulsion between diquarks if the diquark-diquark coupling constant is higher than a critical value $\lambda_C=7.65$. Considering such diquark-diquark repulsion, a positive pressure with no significant variation along the whole strongly interacting region is obtained. A consequence of the diquark-diquark repulsion is that the system maintains its BCS character in the whole strongly interacting region.

\end{abstract}


\maketitle

\section{Introduction}

Shortly after the discovery of asymptotic freedom in QCD \cite{AF}, it was noted \cite{Collins} that the superdense matter might consist of weakly interacting quarks rather than of hadrons. Asymptotic freedom implies that at very high baryon density, QCD is amenable to perturbative techniques \cite{Perturbative}, and since the cores of neutron stars are formed by superdense matter, possible applications of those results to astrophysics \cite{Collins} were envisioned. Later on, however, it was understood that the ground state of the superdense quark system, a Fermi liquid of weakly interacting quarks, is unstable with respect to the formation of diquark condensates \cite{CS-1}, a non-perturbative phenomenon essentially equivalent to the Cooper instability of BCS superconductivity. In QCD, one gluon exchange between two quarks is attractive in the color-antitriplet channel. Thus, at sufficiently high density and sufficiently small temperature $T$, quarks should condense into Cooper pairs, which are color antitriplets. These color condensates break the SU(3) color gauge symmetry of the ground state producing a color superconductor. In the late 90's the interest in color superconductivity (CS) was regained once it was shown, on the basis of different effective theories for low energy QCD \cite{CS-2}, that a color-breaking diquark condensate of much larger magnitude than originally thought may exist already at relatively moderate densities (of the order of a few times the nuclear matter density) and therefore it might be realized in compact stars. At densities much higher than the masses of the u, d, and s quarks, one can assume the three quarks as massless. In this asymptotic region the favored state results to be the so-called color-flavor-locking (CFL) state \cite{CS-2}, characterized by a spin-zero diquark condensate antisymmetric in both color and flavor.

Nevertheless, this picture breaks down at intermediate densities due to the mismatch between the Fermi momenta of different quarks produced by the strange quark mass $M_s$ and the constraints imposed by electric and color neutralities \cite{gCFL}. That is, although the validity of the CFL phase at asymptotically large densities is well established, the next phase down in density is still a puzzle since as a consequence of the pairings with mismatched Fermi surfaces the phase exhibits chromomagnetic instabilities \cite{Chromomagnt-inst}.

One possible scenario where this instability can be avoided occurs if in the region of moderate-low densities the strong coupling constant becomes sufficiently high ($G_D\approx G_S\approx 1/\Lambda^2$, with $G_S$ and $G_S$ denoting the diquark and quark-antiquark coupling constants respectively) \cite{Strong-Coupling, Strong-Coupling-2}. On the other hand, the increase of the coupling constant strength at low density can modify the properties of the ground state as indicated by the significant decrease of the Cooper-pair coherence length, which can reach values of the order of the inter-quark spacing \cite{Coher-length}. As already found in other physical contexts \cite{BCS-BEC}, this fact strongly suggests the possibility of a crossover from a color-superconducting BCS dynamics to a BEC one \cite{R-BCS-BEC-1}-\cite{Baym}, where although the symmetry breaking order parameter (the diquark condensate) is the same, the quasiparticle spectra in the two regions are completely different. As we will show by numerical calculations, in the BCS region, where the diquark coupling is relatively weak, the energy spectrum of the excitations has a fermionic nature, while in the strong-coupling region, formed by the BEC molecules, the energy spectrum of the quasiparticles is bosonic.

As mentioned above, the combination of high densities and relatively low temperatures could exist in the dense cores of compact stars. The cores of neutron star remnants from supernovae collapse have densities several times larger than the saturation density of nuclear matter and temperatures several orders smaller than the superconducting gap. Under these conditions diquark pairs can form and resist the evaporation due to thermal effects. Then, it is natural to ask if a BEC of diquark pairs can take place at the moderately high density that the inner core of neutron stars can reach.

In this paper, we will show, through the mean-field analysis of the equation of state (EoS) of a simple system with four-fermion interactions, that by increasing the diquark coupling strength the matter pressure decreases up to negative values once the crossover from the BCS region to the BEC one takes place. This result would hint that if the density decreases to values where the coupling becomes sufficiently strong, the matter pressure turns to be negative and the system becomes unstable under the effect of gravity. Nevertheless, this is a naive picture that ignores the diquark-diquark interactions. As it was pointed out in Ref. \cite{Wilczek}, together with the fact that there exists an attractive channel between quarks that favors the diquark formation, there is, as a corollary, a diquark-diquark repulsion. This repulsion is due to the cross-channel unfavorable correlations between the quarks belonging to different diquarks. Hence, when the diquark repulsion is self-consistently taken into account in the EoS of this system, the instability previously found in the strong coupling region is removed, and the pressure is stabilized with no significant variation through out that region. The increase of the diquark repulsion, which is produced by the raise of the energy gap in the strong coupling region, compensates the effect of the decay of the chemical potential, that as known, makes an important contribution to the EoS \cite{Alford}. Yet, the price of the stabilizing effect produced by the diquark repulsion is that the Bose-Einstein nature of the strongly interacting system under study is lost, as we will discuss below. Our finding is calling attention on one hand to the necessity of including the diquark-diquark repulsive potential in the studies of the BCS-BEC crossover of strongly interactions, something that has been ignored up to now in previous works, and on the other hand, it is indicating that to include the diquark-diquark repulsion can completely change the understanding of this phenomenon in the context of strong interactions.

\section{BCS-BEC crossover and quasiparticle spectrum}
Our main goal in this section is to determine through a numerical analysis the threshold value of the attractive coupling constant between quarks that marks the crossover from the BCS to the BEC region in the frame of the simple model under consideration.

For our investigation, we consider a simplified pure fermion system represented by the four-fermion interaction Lagrangian density of Ref. \cite{R-BCS-BEC-1}
\begin{equation}\label{lagrangian}
\mathcal{L}=\bar{\psi}(i\gamma^{\mu}\partial_{\mu}+\gamma_0\mu-m)\psi+\frac{g}{4}(\bar{\psi}i\gamma_{5}C\bar{\psi}^T)(\psi^TCi\gamma_5\psi)
\end{equation}
In (\ref{lagrangian}), $C=i\gamma_0\gamma_2$ is the charge conjugation matrix, $m$ the fermion mass, $\mu$ the chemical potential defining the Fermi energy, and $g$ the attractive coupling constant in the $J^P=0^+$ channel that parameterizes the strength of the interaction. Varying the strength of $g$ yields the crossover from BCS (for a weak $g$) to BEC (for a strong $g$). The results we will obtain should not qualitatively change when additional internal fermion degrees of freedom, other than spin, are considered. It is due to the fact that the essence of the phenomenon under investigation is uniquely related to the change in the nature of the spectrum of the quasiparticles, which is determined by the variation of the  diquark-pair binding energy as a function of the coupling constant strength.

After introducing the Hubbard-Stratonovich transformation with gap parameter $\Delta=\langle g\psi^TCi\gamma_5\psi/2\rangle$, we have that the system free energy at finite temperature in the mean-field approximation is
\begin{equation}\label{MF-potential}
\Omega_T=-\frac{1}{\beta}\sum_{n=0}^\infty\int\frac{d^3k}{(2\pi)^4}~Tr \ln~[\beta G^{-1}(i\omega_n,\textbf{k})]+\frac{\Delta^{2}}{g},
\end{equation}
where $G^{-1}(i\omega_n,\textbf{k})$ is the inverse propagator in Nambu-Gor'kov space in the field basis $\Psi^T=(\psi, \psi_C)$, with $\psi_C=C\overline{\psi}^T$ being the charge-conjugate spinors,
\begin{equation}\label{Propagator}
G^{-1}(i\omega_n,\textbf{k})=(\omega_n +\mu \sigma_3)\gamma_0-\gamma \cdot \textbf{k}-m+i\gamma_5\Delta \sigma_+ +i\gamma_5 \Delta^*\sigma_-
\end{equation}
Here, $\omega_n=(2n+1)\pi/\beta$ are the fermion Matsubara frequencies, and $\sigma_{\pm}=\sigma_1\pm i\sigma_2$, with $\sigma_{1,2}$ denoting the corresponding Pauli matrices. After taking the trace and the sum in Matsubara frequencies in (\ref{MF-potential}) it is obtained in the zero-temperature limit
\begin{equation}\label{T0-potential}
\Omega_{0}=-\sum\limits_{e=\pm 1}\int_{\Lambda}\frac{d^{3}k}{(2\pi)^{3}}\;\epsilon_{k}^{e}+\frac{\Delta^{2}}{g},
\end{equation}
where $\Lambda$ is an appropriate momentum cutoff to regularize the momentum integral in the ultraviolet, and the quasiparticle energy spectrum, $\epsilon_{k}^{e}$, is given by
\begin{equation}\label{spectrum}
\epsilon_{k}^{e}=\sqrt{(\epsilon_{k}-e\mu)^{2}+\Delta^{2}}, ~~~~ \epsilon_{k}=\sqrt{k^{2}+m^{2}}, ~~~~ e=\pm.
\end{equation}
The spectra corresponding to different $e=\pm$ values denote the particle ($e=+$) and
 antiparticle ($e=-$) contributions.
\begin{figure}
\includegraphics[width=0.3\linewidth, angle=-90]{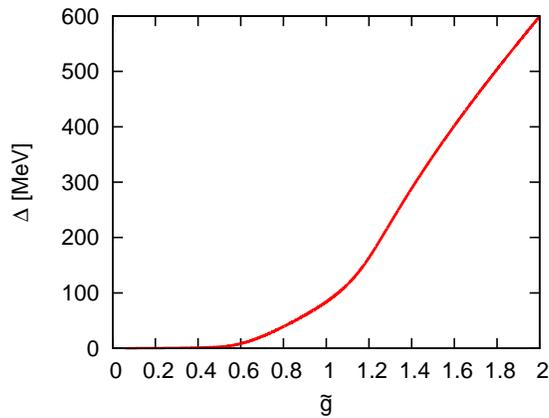}
\caption{Energy gap, $\Delta$, vs the coupling constant $\widetilde{g}=g\Lambda^2$ for a free-diquark gas.}
\label{fig:mudelta}
\end{figure}

\begin{figure}
\includegraphics[width=0.3\linewidth, angle=-90]{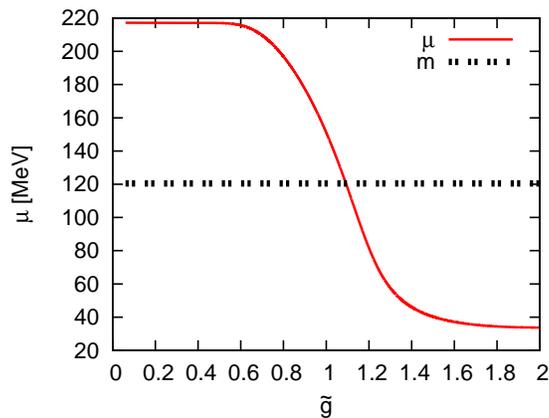}
\caption{Chemical potential, $\mu$, and mass $m$ vs $\widetilde{g}=g\Lambda^2$ for a free-diquark gas.}
\label{fig:mu}
\end{figure}

\begin{figure}
\includegraphics[width=0.3\linewidth, angle=-90]{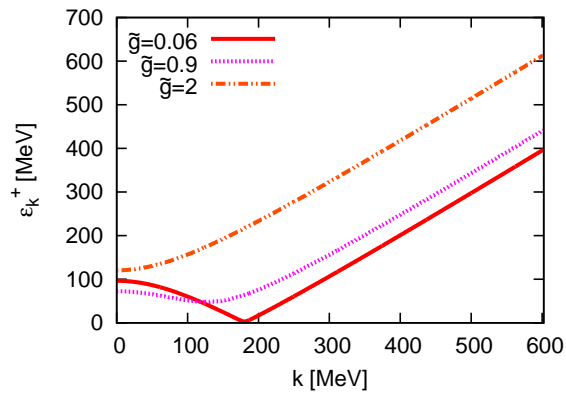}
\caption{$\epsilon_{k}^{+}$ vs $k$ plotted for different $\widetilde{g}=g\Lambda^2$ values.}
\label{fig:spectrum}
\end{figure}

A stable phase must minimize the free energy with respect to the variation of the gap parameter $\partial\Omega_{0} /\partial \Delta = 0$. Then, from (\ref{T0-potential}) we obtain the gap equation
\begin{equation}\label{gap}
1=g\int_{\Lambda}\frac{d^{3}k}{(2\pi)^{3}}\left[\frac{1}{2\epsilon_{k}^{+}}+\frac{1}{2\epsilon_{k}^{-}}\right]
\end{equation}

As usual in the study of the BCS-BEC crossover we will consider a canonical ensemble where the particle number density, $n_{F}=-\partial\Omega_{0}/\partial\mu$, is fixed through the Fermi momentum $P_F$ as $n_{F}=P_{F}^{3}/{3\pi^{2}}$. Then, from (\ref{T0-potential}) we get
\begin{equation}\label{neutrality}
\frac{P_{F}^{3}}{3\pi^{2}}=-\int_{\Lambda}\frac{d^{3}k}{(2\pi)^{3}}\left[\frac{\xi_{k}^{+}}{\epsilon_{k}^{+}}-\frac{\xi_{k}^{-}}{\epsilon_{k}^{-}}\right]
\end{equation}
with
\begin{equation}
\xi_{k}^{\pm}=\epsilon_{k}\mp\mu
\end{equation}

Now, we solve numerically the system of Eqs.~(\ref{gap})~and~(\ref{neutrality}) to find the gap $\Delta$ and chemical potential $\mu$, which correspond to different values of the coupling constant $g$. We scale the theory parameters so to guarantee a relativistic regime that simulates a quark gas at moderate densities. That is, $P_{F}/{\Lambda}=0.3, ~ m/{\Lambda}=0.2$. The results for $\Delta$ and $\mu$ as functions of $g$, in the interval $0.06>\widetilde{g}>2$, with $\widetilde{g}=g\Lambda^2$, and for $\Lambda=602.3$ MeV, are shown in Figs. 1 and 2 respectively. From Fig.1 we see that increasing the coupling strength, the energy gap $\Delta$ becomes larger. This suggests that the binding energy of the diquark condensate approaches that of a Bose-Einstein condensate (with a smaller coherence length $\xi\sim 1/\Delta$) at stronger coupling. To corroborate that this is the case, we should observe the behavior of the chemical potential with increasing the coupling strength in Fig. 2.

As known, the condition $\mu < m$ is characteristic of a relativistic Bose gas \cite{BEC-Rel}. From Fig. 2 we see that for this simple model there exists a critical value for the coupling constant $\widetilde{g}_{cr}\sim 1.1$ beyond which the condition $\mu < m$ is satisfied. Hence, the quasiparticle spectrum corresponding to coupling constants smaller and larger than $\widetilde{g}_{cr}$ should correspond to fermion-like and boson-like behaviors, respectively. In Fig. 3, we have plotted the quasiparticle spectra, $\epsilon_k^+$, corresponding to different values of the coupling constant. The gap, $\Delta$, and chemical potential, $\mu$, entering in the quasiparticle spectrum (\ref{spectrum}) are obtained as solutions of Eqs. (\ref{gap}) and (\ref{neutrality}) for each value of $g$. From their graphical representations in Fig. 3, we can see that for the spectra corresponding to $\widetilde{g}=0.06$ and $0.9$ the minimum of their dispersion relations
occurs at $k=\sqrt{\mu^{2}-m^{2}}$, with
excitation energy given by the gap $\Delta$, a behavior
characteristic of quasiparticles in the BCS regime. On the other hand, for $\widetilde{g}=2$, the minimum of the corresponding spectrum occurs at $k=0$, with excitation energy $\sqrt{(\mu-m)^{2}+\Delta^{2}}$, which is typical of Bosonic-like quasiparticle. Therefore, it is corroborated that $\widetilde{g}_{cr}$ is the threshold value for the BCS-BEC crossover in this model. In other words, for $\widetilde{g}<\widetilde{g}_{cr}$, we have $\mu > m$ and the quasiparticles exhibit fermionic-like modes, while for $\widetilde{g}>\widetilde{g}_{cr}$, we have $\mu < m$ and the quasiparticles are characterized by bosonic-like modes.

We should underline that the obtained value of the critical coupling for the BCS-BEC crossover ($\widetilde{g}_{cr}\sim 1.1$) is in the allowed range of values for the strong coupling regime of QCD, where $\widetilde{g}=g\Lambda^2$ is expected to be of the order of the quark-antiquark coupling $G_S\Lambda^2$ \cite{Strong-Coupling}. As found in Ref. \cite{Klevansky}, once fixed the up and down quark masses with equal values $m_{u,d}=5.5 MeV$, the four observables of vacuum QCD with values, $m_{\pi}=135.0$ MeV, $m_K=497.7$ MeV, $m_{\eta'}=957.8$ MeV and $f_{\pi}=92.4$ MeV, are obtained for $\Lambda=602.3$ MeV (which is the value used in our calculations) and $G_S\Lambda^2=1.835$.

\section{Unstable BEC free-diquark region}

To find the system EoS it is needed to find the system energy density and pressure. In the case we are investigating those magnitudes will depend on the coupling-constant strength. Therefore, varying the values of $g$ from $g<g_{cr}$ to $g>g_{cr}$ we will be able to describe the EoS corresponding to the BCS and BEC regimes respectively.

The energy density and pressure are obtained respectively from the $\langle T_{00}\rangle$ and $\langle T_{ii}\rangle$ components of the quantum-statistical average of the energy momentum tensor. For an isotropic system, as the one we are considering, the covariant structure of the $\langle T_{\mu \nu}\rangle$ tensor is given as \cite{Ferrer}
\begin{equation}\label{energy-momentum}
\frac{T}{V}~\langle T_{\mu \nu}\rangle=(\Omega_{0}+B)g_{\mu \nu}+(\mu n_{F}+TS)u_\mu u_\nu
\end{equation}
where $V$ is the system volume, $T$ the absolute temperature, $S$ the entropy, and $u_\mu$ the medium 4-velocity with value $u_\mu=(1,\overrightarrow{0})$ in the rest frame. In (\ref{energy-momentum}) we introduced the bag constant $B$ to account for the energy difference between the perturbative vacuum and the true one. In that way, we are modeling what occurs in the case of quark matter, where the asymptotically-free phase of quarks forms a perturbative regime (inside a bag) which is immersed in the nonperturbative vacuum. This scenario is what is called the MIT bag model \cite{MIT}. The creation of the bag costs free energy. Then, in the energy density, the energy difference between the perturbative vaccum and the true one should be added. Essentially, that is the bag constant $B$ characterizing a constant energy per unit volume associated to the region where the quarks live. From the point of view of the pressure, $B$ can be interpreted as an inward pressure needed to confine the quarks into the bag. In the numerical calculations we will take $B^{1/4}=145$ MeV, which is a value compatible with that found in the MIT model.

Then, the energy density and pressure of the system in the zero-temperature limit are respectively calculated from
\begin{equation}\label{eq:ep}
\varepsilon=\Omega_{0}+\mu n_{F}+B, ~~~~ p=-\Omega_{0}-B
\end{equation}

The results for $\varepsilon$ and $p$ are plotted in Fig. 4 versus the coupling-constant strength $\widetilde{g}$. There, we can see that the system energy density is increasing with the coupling strength, while the pressure is decreasing up to get negative values at coupling constants corresponding to the BEC regime. The appearance of a negative pressure for the diquark free gas in the BEC region indicates that the free-diquark system is unstable.

\begin{figure}
\includegraphics[width=0.3\linewidth, angle=-90]{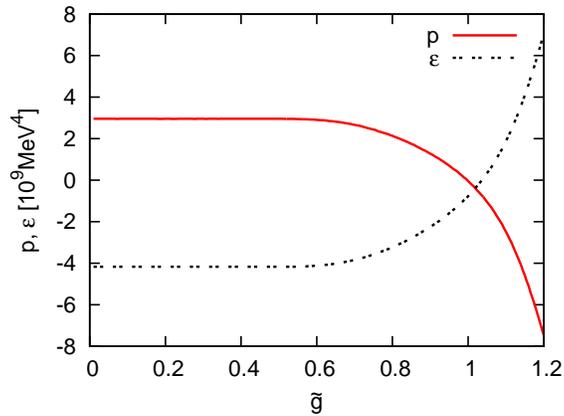}
\caption{Energy density, $\varepsilon$, and pressure, $p$, vs the coupling strength $\widetilde{g}=g\Lambda^2$ for a free-diquark gas.}
\label{fig:pressureenergy}
\end{figure}

\section{Equation of State of self-interacting diquarks}

The system pressure decay in the BEC region obtained in Fig. 4 is an expected result since the absence of repulsion between the diquarks makes their Bose-Einstein condensation inevitable with the corresponding decrease of the matter pressure. Nevertheless, as we will show in this section, the contribution of the diquark-diquark repulsion in the EoS of the strongly interacting system compensates the decreasing tendency due to the Bose-Einstein condensation and consequently rendering a constant pressure throughout the strongly interacting region.

The modeling of self-interacting diquarks in the context of a $\phi^4$ boson theory was initially developed in \cite{Diquarks} and then applied to different situations in \cite{Diquark Appl.}. In our case, it can be achieved by introducing a $\lambda \Delta^4$ term in the free energy (\ref{MF-potential})
\begin{equation}\label{MF-potential-2}
\Omega_T=-\frac{1}{\beta}\sum_{n=0}^\infty\int\frac{d^3k}{(2\pi)^4}~Tr \ln~[\beta G^{-1}(i\omega_n,\textbf{k})]+\frac{\Delta^{2}}{g}+\lambda\Delta^4,
\end{equation}

Hence, the system energy density and pressure given in (\ref{eq:ep}) become
\begin{equation}\label{eq:ep-2}
\varepsilon=\Omega_{0}+\lambda\Delta^4+\mu n_{F}+B, ~~~~ p=-\Omega_{0}-\lambda\Delta^4-B
\end{equation}

A possible value for the coupling constant $\lambda$ was estimated as $\lambda=27.8$ in \cite{Diquarks}. It was found taking into account the quark interactions in the context of a modified P-matrix formalism of Jaffe and Low \cite{Jaffe}.

The values for $\Delta$ and $\mu$ obtained for $\lambda=27.8$ from the modified gap equation after including the diquark-diquark repulsive interaction term
\begin{equation}\label{gap-2}
1=g\int_{\Lambda}\frac{d^{3}k}{(2\pi)^{3}}\left[\frac{1}{2\epsilon_{k}^{+}}+\frac{1}{2\epsilon_{k}^{-}}\right]-2\lambda g \Delta^2
\end{equation}
and (\ref{neutrality}), are given in Fig. 5.

\begin{figure}
\includegraphics[width=0.3\linewidth, angle=-90]{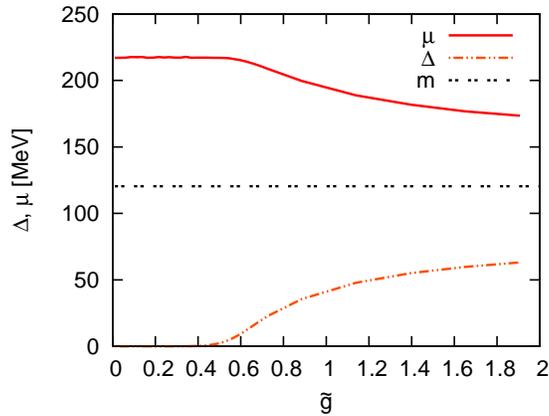}
\caption{Energy gap, $\Delta$, chemical potential $\mu$ and mass $m$ vs the coupling constant $\widetilde{g}=g\Lambda^2$ for a self-interacting diquark gas with $\lambda=27.8$.}
\label{fig:pressureenergy}
\end{figure}

\begin{figure}
\includegraphics[width=0.3\linewidth, angle=-90]{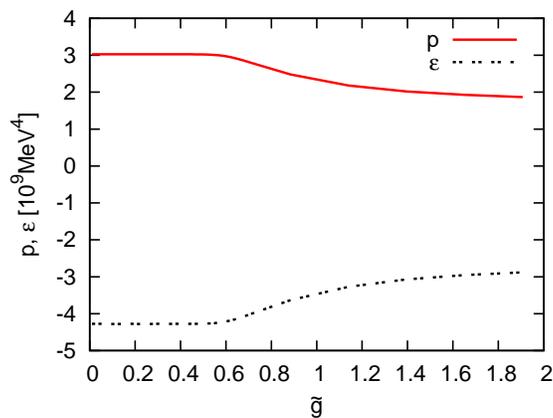}
\caption{Energy density, $\varepsilon$, and pressure, $p$, vs the coupling strength $\widetilde{g}=g\Lambda^2$ for a self-interacting diquark gas with $\lambda=27.8$.}
\label{fig:pressure-vs-energy}
\end{figure}

The repulsive interaction between diquarks makes a significant contribution to the energy density and pressure (\ref{eq:ep-2}) as can be seen comparing Figs. 4 and 6. From Fig. 6, it is apparent that the instability produced by a negative pressure in the BEC region disappears. The matter pressure now remains almost the same in the whole strongly interacting region. In this scenario, the repulsion between diquarks produce enough outward pressure to elude the star collapse. On the other hand, this same effect prevents the gas condensation into a zero momentum ground state at zero temperature. The fact that the diquark repulsion prevails against the Bose-Einstein condensation is reflected in the behavior of the system chemical potential that now can never cross the constant $m$ line in Fig. 5. As we checked numerically, once the diquark repulsion is taken into account, the condition $\mu < m$ is never reached. Hence, the BCS phase is maintained for the whole interaction range.

If one considers arbitrary values of $\lambda$, one can show that if the repulsion is weak enough ($\lambda<\lambda_c=7.65$), the BEC dynamics can be reached by increasing $g$, and consequently, the system pressure becomes negative. On the contrary, if $\lambda>\lambda_c$, the pressure is positive for the whole range of considered g-values. But in this last case one has $\mu>m$ for all those g-values, implying the absence of a BEC region. Notice that $\lambda_c$ is smaller than the estimated value $\lambda=27.8$ \cite{Diquarks}. In conclusion, we find that there is no way to put together a BEC dynamics with a positive pressure; meaning that a gravitational-bound compact star cannot be formed by BEC quark molecules.

We should highlight that our approach is different from that developed in Ref. \cite{Diquarks}. In our case, the diquark repulsion effect is treated in the EoS as a dynamical variable, which is determined through the gap equation (\ref{gap-2}) and the number density constraint (\ref{neutrality}); while in \cite{Diquarks} the contribution of the diquark repulsive potential to the energy density and pressure was modeled by assuming an ad hoc Gaussian diquark distribution for the occupation of the states with momenta $k>0$. In this way, it was prevented that at zero temperature the ground state of the diquark gas condensed in the $k=0$ state characteristic of the Bose-Einstein condensation phenomenon of a free Boson gas.

\section{Summary and Final Remarks}

The goal of this paper is to illustrate the behavior of a diquark gas in the strong coupling regime. We started by considering a free diquark system. In this system we found that as the strength of the attractive coupling between quarks increases, the chemical potential transits from being larger than the quark mass to being smaller, an indication of a crossover from the BCS region to the BEC one. A consequence of the crossover to BEC is that the matter pressure decays to zero, and even reaches negative values, a sign that the BEC regime cannot be realized in the interior of a neutron star.

We then considered whether this instability could be removed by the introduction of a repulsive force between diquarks. In this case, the pressure collapse can be prevented, since a sufficiently strong ($\lambda>\lambda_c$) diquark-diquark repulsion will hinder the overlapping of the diquarks in the ground state. But the implications of reaching a stable state in the strong diquark coupling regime is that the system maintains its BCS nature for the whole $g$-value range.
As shown then by numerical calculations for $\lambda=27.8>\lambda_c$ in particular, the contribution of the diquark self-interaction is sufficient for stabilizing the system that then acquires an EoS stiff enough (see Figs. 6 and 7) to prevent the collapse that would be caused by the formation of a pressureless gas of diquark molecules in the BEC region. The fact that the diquark negative pressure prevents the formation of a Bose-Einstein diquark condensate at zero temperature in the strongly interacting region is substantiated by the condition $\mu > m$ for the whole interacting region. Thus, a self-interacting diquark system will not form a Bose gas.

The inclusion of the diquark-diquark negative presure in the strongly interacting system makes the EoS stiffer, as can be corroborated from our results. This effect can be important to accommodate quark matter into the EoS of neutron stars of high masses, as for instance PSR J1614-2230 \cite{Hessels} with an inferred value of $1.97\pm0.04 M_\odot$ \cite{Demorest}. As known, when quark, or other degrees of freedom like hyperons or bosons are considered, the corresponding EoS softens and it cannot support highly massive compact stars \cite{Lattimer}, unless in the case of quarks, if there exists a color superconducting phase with strong interactions between the quarks \cite{Ozel}. Nevertheless, the diquark-diquark repulsion has not been considered in previous approaches, and its effects are worth to be investigated.

\begin{figure}
\includegraphics[width=0.3\linewidth, angle=-90]{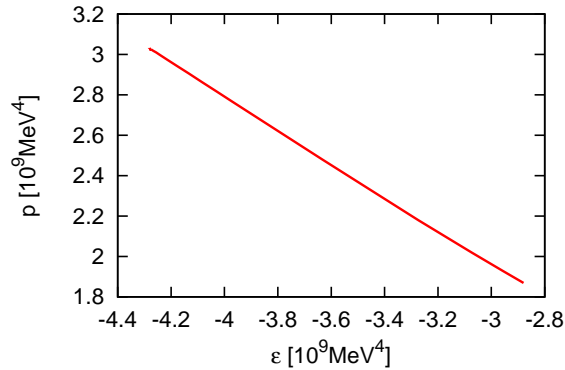}
\caption{Equation of state of the strongly interacting system with diquark-diquark repulsive interaction ($\lambda=27.8$) for different values of the coupling strength $\widetilde{g}$ around the critical value $\widetilde{g}_{cr}$.}
\label{fig:pressure-vs-energy}
\end{figure}

Nevertheless, we should mention that stars formed by bosons (the so-called boson stars) have been theoretically considered since long ago starting with Wheeler's notion of geons \cite{Wheeler} (see Ref. \cite{B-stars} for recent reviews). A peculiarity of those stars is that the internal degrees of freedom are bosonic, so the mechanism to stabilize the star against its self-gravity cannot be through the fermionic degeneracy pressure, but by the limitations imposed by the Heisenberg uncertainty principle for the mass and radius of the star.

The result we are reporting in the frame of the simple model of Eq. (\ref{lagrangian}) should be investigated in more realistic models as those of Ref. \cite{Strong-Coupling, Strong-Coupling-2} for strong-coupling regimes. Nevertheless, we expect that the main outcome of this paper will remain valid. That is, the system pressure will be stabilized by the diquark repulsion. Our expectation is based on the fact that by increasing the coupling strength the gap will increase so as to make significant the contribution of the outward pressure associated with the diquark repulsive force. A nontrivial problem that remains unsolved is to develop the diquark-diquark interaction from first principles. We envision that it will require to start from an extended effective theory with a higher number of fermion interactions (as for example, the eight-fermion interaction model introduced in Ref. \cite{Osipov}) that can give rise in principle to a self-interaction term between the diquark condensates. Such a study, however, is out of the scope of the present work.

An interesting question to be studied in this scenario is the possible effect of an applied magnetic field. As already estimated in Ref. \cite{Ferrer}, magnetic fields of order $10^{19}-10^{20}$ G can coexist in the core of neutron stars with quark matter. By increasing the magnetic field strength, $\Delta$ increases \cite{MCFL}, and the system is lead to crossover from the BEC region to the BCS one \cite{Wang}. At very strong magnetic fields, when all the particles are localized in the lowest Landau level, only BCS diquarks are allowed \cite{Wang}. On the other hand, the pure magnetic contribution to the pressure is negative \cite{Laura}. Thus, the magnetic field will have a double effect in the pressure whose consequences should be elucidated in the frame of the strongly interacting system.

In those regions of relatively low densities there is of course the possibility that the repulsion between the diquarks catalyzes a phase transition to other ground state configurations such as a hadronic phase with a well identified fermion nature able to produce the degeneracy pressure needed to compensate for the gravitational pull. Other possibilities to be investigated are the viability, through their EoS, of some inhomogeneous phases, such as those formed by density waves \cite{D-W}, quarkyonic chiral spirals \cite{C-S}, inhomogeneous Fulde-Ferrel state \cite{Sedrakian} or quark clusters in solid or liquid states \cite{Renxiu}, that can in principle be realized in quark matter at moderate density.


\begin{acknowledgments}
The work of E. J. Ferrer has been supported in part by DOE Nuclear Theory grant DE-SC0002179 and the work of
J. Keith by UTEP-COURI-2011 grant. The authors thank V. de la Incera for enlightening discussions and comments.

\end{acknowledgments}

\end{document}